\documentclass[11pt]{article}
\usepackage{graphicx,amsmath,amsthm,mathrsfs,amssymb, braket, verbatim}
\usepackage[usenames]{color}
\usepackage{ulem,cite}
\usepackage{authblk}
\usepackage{tcolorbox}

\newcommand{\ee}{\end{equation}}
\newcommand{\word}[1]{\,\,\mbox{#1}\,\,}
\newcommand{\reff}[1]{(\ref{#1})}
\newcommand{\beq}{\begin{equation}}
\newcommand{\eeq}[1]{\label{#1}\end{equation}}
\newcommand{\beqa}{\begin{eqnarray}}
\newcommand{\eea}{\end{eqnarray}}
\newcommand{\eeqa}[1]{\label{#1}\end{eqnarray}}
\newcommand{\beg}{\begin{equation*}}
\newcommand{\eeg}{\end{equation*}}

\newcommand{\abs}[1]{\lvert#1\rvert}

\newcommand{\bsplit}{\begin{split}}
\newcommand{\esplit}{\end{split}}

\usepackage{subfig} 
\usepackage{circuitikz} 
\usepackage[capposition=bottom]{floatrow} 
\usepackage{epigraph} 
\usepackage{appendix}
\usepackage{rotating}
\allowdisplaybreaks

\title{First and second-order relativistic corrections to the two and higher-dimensional isotropic harmonic oscillator \\obeying the spinless Salpeter equation}

\author[]{Ariel Edery\thanks{aedery@ubishops.ca}}
\author[]{Philippe Laporte\thanks{plaporte13@ubishops.ca}}

\affil[]{Department of Physics and Astronomy, Bishop's University, 2600 College Street, Sherbrooke, Qu\'{e}bec, Canada, J1M 1Z7.\vspace{1em}}

\begin{document}
\date{}
\maketitle
\begin{abstract}
We study the relativistic version of the $d$-dimensional isotropic quantum harmonic oscillator based on the spinless Salpeter equation. This has no exact analytical solutions. We use perturbation theory to obtain compact formulas for the first and second-order relativistic corrections; they are expressed in terms of two quantum numbers and the spatial dimension $d$. The formula for the first-order correction is obtained using two different methods and we illustrate how this correction splits the original energy into a number of distinct levels each with their own degeneracy. Previous authors obtained results in one and three dimensions and our general formulas reduce to them when $d=1$ and $d=3$ respectively. Our two-dimensional results are novel and we provide an example that illustrates why two dimensions is of physical interest. We also obtain results for the two-dimensional case using a completely independent method that employs ladder operators in polar coordinates. In total, three methods are used in this work and the results all agree.  
\end{abstract}
\setcounter{page}{1}
\newpage
\section{Introduction}\label{Intro}
The harmonic oscillator plays a crucial role in quantum mechanics and quantum field theory. In quantum mechanics, the quadratic potential is one of the few potentials that yields exact analytical results. Moreover, the result is somewhat generic since any smooth potential that possesses a minimum can be approximated as a quadratic potential in the neighborhood of its equilibrium point. The energy of the ground state of the quantum harmonic oscillator (QHO) is not zero, a fact which takes on great importance in quantum field theory and in cosmology. A free quantized field like the electromagnetic field can be interpreted as an infinite collection of QHOs \cite{QFT} leading to a vacuum that possesses an infinite energy (or very large depending on the momentum cut-off). This is at the heart of one of the most infamous problems in theoretical physics: the cosmological constant problem \cite{Zel}.  

Due to the importance of the QHO, many authors have studied relativistic versions. As pointed out in \cite{Nagiyev}, relativistic generalizations of the QHO are not unique. The situation is different from the case of the Coulomb potential, which can be recovered from the tree level Feynman diagram between two charged particles in quantum electrodynamics (see \cite{Zee} for a derivation of the Coulomb potential in quantum field theory). For the QHO, one can define a Dirac oscillator for spin 1/2 particles and a Klein-Gordon oscillator for spinless particles both of which have been studied by many authors \cite{Moshinsky, Lucha1, Lucha2, Znojil, Poszwa}. An interesting model called finite-difference relativistic quantum mechanics was solved in \cite{Nagiyev}. In that relativistic model, the authors were able to obtain exact analytical results for the energies of the relativistic three dimensional isotropic singular oscillator. 

In this paper, we use a model of the relativistic QHO based on the spinless Salpeter equation \cite{Salpeter} as it is the most straightforward relativistic generalization. The spinless Salpeter equation can be viewed as an approximation to the Bethe-Salpeter formalism \cite{Bethe} within relativistic quantum field theories where the interactions are instantaneous and the spin degrees of freedom are neglected. The non-relativistic kinetic term $\frac{p^2}{2m}$ is simply replaced by its relativistic counterpart $\sqrt{m^2\,c^4 + p^2\,c^2}$ and interactions are described by a coordinate-dependent potential $V(x)$. This equation has previously been applied to describe bound-state constituents. For example, for two particles of equal mass $m$ in a bound state, the Hamiltonian $H$ in the center-of-momentum frame of these constituents is given by $H = 2 \sqrt{p^2\,c^2 + m^2\,c^4} + V(x)$ where $p$ is the relative momentum and $x$ is the relative coordinate. This Hamiltonian has been used for the semi-relativistic description of hadrons as bound states of quarks within the framework of potential models \cite{Grome,Lucha3, Lucha4}. Because of the difficulty in handling the square root in the Hamiltonian, many authors have also studied the spinless Salpeter equation numerically \cite{Durand,Jacobs,Gara,Rup,Chen,Fulcher}.
 
In our work, we consider a single particle moving in an isotropic harmonic potential. The ``Salpeter" Hamiltonian for the relativistic QHO is then given by 
\beq
H= \sqrt{m^2\,c^4 + p^2\,c^2} -m\,c^2 + \dfrac{1}{2} m\, \omega^2\,r^2
\eeq{Intro}
where  $p$ is the momentum operator, $r$ the radius, $m$ the mass of the particle, $c$ the speed of light and $\omega$ a free parameter (we subtracted out the rest mass of the particle as it is convenient to set the kinetic term to zero when $p=0$). In contrast to the non-relativistic case, there are no known exact analytical results for the above Hamiltonian though lower and upper energy bounds have been obtained in three dimensions \cite{Lucha1, Lucha2}. Authors have therefore turned to perturbation theory to obtain relativistic corrections. Expanding the above Hamiltonian operator in powers of $\frac{p^2}{m^2\,c^2}$ one obtains
\beq
\begin{split}
H &= \dfrac{p^2}{2\,m}+ \dfrac{1}{2} m\, \omega^2\,r^2-\dfrac{1}{8}\dfrac{p^4}{m^3\,c^2} +\dfrac{1}{16}\dfrac{p^6}{m^5\,c^4} + ...\\
&= H_0 +H_1' +H_2'+...
\end{split}
\eeq{Pert}
where 
\beq
H_0=\frac{p^2}{2\,m} + \frac{1}{2}m\, \omega^2\,r^2
\eeq{NR}
is the non-relativistic Hamiltonian and 
\beq
H_1'= -\dfrac{p^4}{8\,m^3\,c^2} \word{and} H_2'=\dfrac{p^6}{16\,m^5\,c^4}
\eeq{rel} 
are first and second-order perturbations to the Hamiltonian $H_0$ respectively. Using perturbation theory, analytical results for the Hamiltonian \reff{Intro} were previously obtained in one and three dimensions: first-order relativistic corrections in three dimensions were obtained in \cite{Znojil} whereas first and second-order relativistic corrections in one and three dimensions were obtained in \cite{Poszwa}. 
 
In this paper we obtain explicit analytical formulas for the first and second order relativistic corrections to the isotropic QHO governed by the Salpeter Hamiltonian \reff{Intro}, that are valid in any spatial dimension $d$. The problem is solved in spherical coordinates and the results are expressed in terms of the dimension $d$ and two quantum numbers $n$ and $\ell$ which are non-negative integers. Our first-order correction formula reduces to that obtained in \cite{Znojil,Poszwa} for $d=1$ and $d=3$, and our second-order correction formula reduces to that obtained in \cite{Poszwa} for $d=1$ and $d=3$. Our $d=2$ results are novel and we also solve the two-dimensional case using a completely independent method that makes use of ladder operators in polar coordinates. The result obtained by substituting $d=2$ into our general formulas matches those obtained using the ladder operator technique. The two-dimensional case is of potential physical interest as we point out that for a particular choice of magnetic field strength, a charged particle moving in a uniform magnetic field 
$\textbf{B}= B_0 \,\hat{z}$ and a linear electric field $\textbf{E}=-k \,z\,\hat{z}$ (where $B_0$ and $k$ are constants) can have quantized energy levels that are equivalent to those of the two-dimensional $isotropic$ QHO.  

The general formula for the first-order correction is obtained in two different ways: one using a $d$-dimensional Kramers-Pasternak type relation that we derive that relates the expectation value of a power of $r$ to the expectation value of other powers of $r$ \cite{Pasternak} (see \cite{Zettili,Griffith} for applications)  and a second using a recurrence relation obeyed by the eigenfunctions when they are multiplied by $r^2$. The two independent methods yield the same result providing a strong confirmation of the general formula. We determine how the first-order relativistic correction splits the original energy into distinct levels each with their own degeneracy.
 
The energy of the (unperturbed) $d$-dimensional isotropic QHO is degenerate (except for $d=1$). Nonetheless, if one works in spherical coordinates one can use non-degenerate perturbation theory to calculate relativistic corrections as one does in the case of the hydrogen atom \cite{Shankar,Zettili,Griffith}. Non-degenerate perturbation theory can be used if one can find a Hermitian operator A that has the following two properties: a) it commutes with $H_0$ and the perturbations $H_1'$ and $H_2'$ and b) the degenerate states of $H_0$ are also eigenstates of A with distinct eigenvalues \cite{Griffith}. Let us now identify this Hermitian operator. Consider the three-dimensional isotropic harmonic oscillator in spherical coordinates. The states $\ket{n \ell m}$ are simultaneous eigenstates of the Hamiltonian operator $H_0$ and the angular momentum operators $L^2$ and $L_z$ with eigenvalues 
$(2\,n +\ell +3/2)\,\hbar \,\omega$, $\hbar^2 \,\ell(\ell+1)$ and $\hbar \,m$ respectively where $n$ and $\ell$ are non-negative integers and $m$ runs from $-\ell$ to $\ell$ inclusively in steps of unity. Consider two eigenstates $\ket{n_1 \ell_1 m_1}$ and $\ket{n_2 \ell_2 m_2}$. They are degenerate if $2n_1+\ell_1 = 2n_2+\ell_2$ with $\ell_1\ne \ell_2$ and/or $m_1 \ne m_2$. If $\ell_1 \ne \ell_2$, then when $L^2$ acts separately on the two degenerate states, the eigenvalues will be distinct and if $m_1\ne m_2$, then when $L_z$ acts separately on the two degenerate states, the eigenvalues will be distinct. Therefore, $L^2$ and/or $L_z$ will have distinct eigenvalues. Moreover, $L^2$ and $L_z$ both commute with $H_0$ and the perturbations $H_1'$ and $H_2'$. Therefore, we have identified Hermitian operators, namely $L^2$ or $L_z$, that have the two required properties; the $\ket{n \ell m}$ basis can therefore be used in nondegenerate perturbation theory. It is easy to see that this result generalizes to higher dimensions where there are more angular quantum numbers. So though obtaining the energies and eigenfunctions for the (unperturbed) $d$-dimensional isotropic QHO is much easier in Cartesian coordinates, those eigenfunctions do not form the correct basis for use in nondegenerate perturbation theory. 

Our paper is organized in the following fashion. In section 2 we discuss the non-relativistic $d$-dimensional isotropic QHO in spherical coordinates. The goal here is to gather results, such as the energy and the eigenfunctions of the unperturbed Hamiltonian $H_0$, in $d$ dimensions. In section 3, we obtain, using two different methods, a general formula for the first-order relativistic correction. The effect of this correction is illustrated in an energy level diagram. In section 4, we make use of ladder operators in polar coordinates to perform an independent calculation of the first and second-order relativistic corrections in two dimensions. We also briefly discuss a physical system that has the same quantized energy levels as the two-dimensional isotropic QHO. In section 5, we obtain a general formula for the second-order relativistic correction valid in $d$ dimensions. The conclusion summarizes our final results and discusses a physical system whose relativistic corrections would be of interest to study in the future.

\section{The $d$-dimensional isotropic QHO in spherical coordinates}\label{QHO}

In this section, we obtain the (unperturbed) energies and eigenfunctions in $d$-dimensional spherical coordinates that are needed to calculate the first-order and second-order relativistic corrections. The radial equation in $d$-dimensions (appendix A) is solved for the potential of the isotropic harmonic oscillator. The main goal of this section is to gather results that are needed elsewhere, leading to a self-contained presentation. The reader interested in the details of higher-dimensional wave equations in spherical coordinates is referred to \cite{Dong} and references therein.
    
\subsection{Energies and eigenfunctions of the $d$-dimensional isotropic harmonic oscillator in spherical coordinates}
The potential of the isotropic harmonic oscillator, valid in any dimension, is given by 
\begin{equation}
	\label{potential}
	V(r)= \frac{1}{2}m\,\omega^2\,r^2 
\end{equation}
and substituting this into the $d$-dimensional radial equation \reff{radial_equation} yields
\begin{equation}
	\label{sandwich}
	u''=\left[-\frac{2m}{\hbar^2}\,E+\frac{m^2\omega^2}{\hbar^2}\,r^2+\frac{\big(d-3+l\,(l+d-2)\big)}{r^2}\right]\,u -\frac{(d-3)}{r}\,u' \,.
\end{equation}
To solve the above equation we first consider two limiting cases. When $r \rightarrow 0$, \reff{sandwich} reduces to 
$$-\frac{\hbar^2}{2m}\,u''-\frac{(d-3)\,\hbar^2}{2\,m\,r}u' +\frac{\left[d-3+l(l+d-2)\right]\,\hbar^2}{2\,m\,r^2}\,u = 0.$$ 
This equation admits two solutions, namely $\alpha_1(r) = r^{l+1}$ and $\alpha_{2}(r) = r^{3 - d - l}$. Since the exponent of $\alpha_2$ is negative when $l>3-d$ it is not a valid solution\footnote{In other words, $\alpha_2$ is not a general solution valid for all possible values of the non-negative integer $l$.} as the eigenfunction would diverge as $r \rightarrow 0$. Thus the valid solution as $r \rightarrow 0$ is $\alpha_1(r) = r^{l+1}$. When $r \rightarrow \infty$, \reff{sandwich} reduces to 
$$-\frac{\hbar^2}{2m}\,u''-\frac{(d-3)\,\hbar^2}{2mr}\,u' +\frac{m\,\omega^2\, r^2}{2}\,u = 0.$$
	This equation admits $\beta_{\pm}(r) = e^{\pm \frac{m\,\omega}{2\,\hbar}r^2}$ as solutions. However, since the solution must converge as $r \rightarrow \infty$, only $\beta_-$ is a valid solution. Thus the valid solution as $r\rightarrow \infty$ is $\beta_-(r) = e^{- \frac{m\,\omega}{2\,\hbar}r^2}$.

It is therefore convenient to express $u(r)$ in the form
\beq 
u(r) = r^{l+1}e^{- \frac{m\omega}{2\hbar}r^2} \,f(r)
\eeq{uf} 
where $f(r)$ is a function to be determined. Substituting the above into \reff{sandwich} yields the following differential equation for $f(r)$:
\begin{equation}
	\label{f}
	f''(r)+\left[2\left(\frac{l+1}{r}-\frac{m\,\omega\,r}{\hbar}\right)+\frac{d-3}{r}\right]f'(r)+\left[-\frac{m\,\omega\,(2l+d)}{\hbar}+\frac{2\,m\,E}
	{\hbar^2}\right] f(r) = 0.
\end{equation}
Consider the power series expansion $f(r)=\sum_{i=0}^\infty a_i\,r^i$, where $a_i$ are coefficients to determine. Substituting this into \reff{f} yields the equation
\begin{equation}
	\label{potato}
	\sum_{i=0}^\infty\left\{(i+1)(i-1+2l+d)\,a_{i+1}\,r^{i-1}+\frac{m}{\hbar}\left(-\omega\,(2i+2l+d)+\frac{2E}{\hbar}\right)\,a_i\,r^i \right\}=0.
\end{equation}
The coefficient for each power of $r$ has to be zero independently. The coefficient of $r^{-1}$ is $(2\,l+d-1)\,a_1$. Therefore $(2\,l+d-1)\,a_1=0$. This implies that $a_1$=0 or $2l+d-1=0$. The latter is possible only when $d=1$ (note that in one dimension $l$ is zero). Therefore for $d\ne 1$, we have $a_1=0$ and for $d=1$ we have $a_1$ free (unconstrained). Equation \reff{potato} leads to the following recurrence relation
\begin{equation}
	\label{recurrence_1}
	(i+2)(i+d+2\,l)\,a_{i+2} +\frac{m}{\hbar}\left(-\omega\,(2\,i+2\,l+d)+\frac{2E}{\hbar}\right)\,a_i=0\text{,     } \forall \,i \in \mathbb{N}_0\,.
\end{equation}
For $d\ne 1$, $a_1 = 0$, and the above yields $a_j=0$ for odd values of $j$; there is a recursion relation only for even coefficients. For $d=1$, one has a separate recursion relation for odd and even coefficients. The power series for $f(r)$ must be truncated or else $f(r) \rightarrow e^{r^2}$ at infinity which would imply that $u(r)$ diverges towards infinity. Therefore, past a certain index $n'$, $a_{n'+2} = 0$. Substituting $i=n'$ in \reff{recurrence_1} yields $\frac{m}{\hbar}\left(-\omega(2\,n'+2\,l+d)+\frac{2E}{\hbar}\right)\,a_{n'}=0$. Therefore the energy of the d-dimensional isotropic QHO is given by
\beq
E_{nl} = \hbar \omega \left(n' + l + \frac{d}{2}\right)
\eeq{Enl1}
where $n'$ is even if $d\ne1$ but can be either odd or even for $d=1$. Letting $n'=2\,n$ the final expression for the energy is 
\beq
E_{nl} = \hbar \omega \left(2\,n + l + \frac{d}{2}\right)\,.
\eeq{Enl} 
For $d\ne1$, $n$ and $l$ are any non-negative integer. For the special case of $d=1$, one sets $l=0$ and $n=N/2$ where $N$ is any non-negative integer. So for $d=1$, the energy is non-degenerate and given by the well known result of $E=(N+1/2)\hbar \omega$ with N=0,1,2,...etc.

We now find the exact expression for $f(r)$. Define a new variable, $\eta(r) = \frac{m\,\omega}{\hbar}r^2$. In terms of this new variable, 
equation \reff{f} reads  
\beq
\eta f''(\eta) + \left[\left(l+\frac{d}{2}-1\right) + 1 - \eta\right]\,f'(\eta) + n\,f(\eta) = 0
\eeq{feta}
where \reff{Enl} was substituted for the energy and the primes denote now derivatives with respect to $\eta$. This corresponds to Laguerre's equation, $xy''+(\alpha + 1 - x)y'+n\,y=0$, with $\alpha = l+\frac{d}{2}-1$ and $x = \eta = \frac{m\omega}{\hbar}\,r^2$. The solution is then  
$f(\eta) = L_n^{\left(l+\frac{d}{2}-1\right)}\big(\eta\big)$, where $L_n^{\alpha}(x)$ are the generalized Laguerre polynomials. Substituting $f(\eta)$ into \reff{uf} yields:  
\begin{equation}
	u_{nl}(\eta) = A_{nl} \,\left(\frac{\hbar}{m\,\omega}\right)^{\left(\frac{l+1}{2}\right)}\, \eta^{\left(\frac{l+1}{2}\right)}\,e^{-\frac{\eta}{2}}\,L_n^{\left(l+\frac{d}{2}-1\right)}(\eta)\,.
\eeq{ueta}
where the normalization constant is given by
\begin{equation}
	\label{A}
	A_{nl} = \left(\frac{m\omega}{\hbar}\right)^{\frac{l}{2}+\frac{d}{4}}\sqrt{\frac{2n!}{\Gamma\left(n+l+\frac{d}{2}\right)}}\,.
\end{equation}   
Except for $d=1$, $n$ and $l$ above are any non-negative integers. For $d=1$ one has to set $l=0$, $n=N/2$ where $N$ is a non-negative integer and replace the radius $r$ by $x \in (-\infty,\infty)$.  

\section{First-order relativistic correction: general formula}\label{1RC}

We now calculate the first-order relativistic correction. We obtain a general formula valid in $d$ dimensions using two different methods. Method I uses a $d$ dimensional Kramers-Pasternak type relation that is derived in appendix B. This relation allows one to obtain the expectation value of a given power of $r$ in terms of the expectation value of other powers of $r$. Method II uses a recurrence relation that is obeyed by the eigenfunctions. The two independent methods yield the same general formula providing confirmation of our result.      

\subsection{First-order relativistic correction: method I}\label{1RC1}

The first-order relativistic correction $E^{(1)}$ is given by the expectation value of $H_1'$: 
\begin{equation}
	\label{e1}
	E^{(1)} =\langle\, H_1'\,\rangle =-\dfrac{1}{8\,m^3\,c^2}\langle\, \textbf{p}^4\,\rangle = -\frac{1}{2mc^2}\left[\,E^2 - 2\,E\,\langle\, V\,\rangle + \langle\, V^2\,\rangle\right]
\end{equation}
where $H_1'$ is given by \reff{rel} and we used the fact that $\textbf{p}^2$ is Hermitian and $\textbf{p}^2\,\psi_n= 2\,m\,(E-V)\psi_n$.
The energy of the $d$-dimensional isotropic QHO (unperturbed) are given by \reff{Enl}
\begin{equation}
	\label{energy}
	E_{nl} = \hbar\,\omega\left(2n+l+\frac{d}{2}\right),
\end{equation}
where $n$ and $l$ are quantum numbers that are non-negative integers and $d$ is the dimension (recall that $d=1$ is an exception with $l=0$ identically and $n=N/2$ where $N$ is a non-negative integer). Applying the Feynman-Hellmann theorem yields $\frac{\partial E_{nl}}{\partial \omega} = \left.\langle\, \frac{\partial H_0}{\partial \omega}\,\rangle\right.$ so that $\left.\langle\, r^2 \,\rangle\right.=\frac{\hbar}{m\,\omega}\left(2n+l+\frac{d}{2}\right)$. Substituting this into \reff{e1} yields
\begin{align}
	E^{(1)} = -\frac{\langle\, V^2\,\rangle}{2\,m\,c^2}= -\frac{m\,\omega^4}{8\,c^2}\left.\langle\, r^4 \,\rangle\right.\,.
\label{ER1}
\end{align}
The quantity $\langle\, r^4 \,\rangle$ is obtained from the $d$-dimensional Kramers-Pasternak type relation derived in appendix B by substituting $s=2$ in \reff{dKramers}: 
\begin{multline}
\frac{m^2\omega^2}{\hbar^2}\,8\,\langle\, r^{4}\,\rangle -\frac{2\,m\,E}{\hbar^2} \,6\,\langle\, r^{2}\,\rangle
+\bigg[4\big(d-3+l\,(l+d-2)\big) + (2-d)(6-d)\bigg]=0\,.
\label{r41}
\end{multline}
The first-order relativistic correction is finally given by
\begin{equation}
	\label{er1}
	\boxed{E^{(1)}=-\frac{\hbar^2\omega^2}{8\,m\,c^2}\Big[6\,n^2 + l^2 +6\,n\,l + 3\,n\,d +l\,d +l + \frac{(d^2+2\,d)}{4}\Big]}\,.
\end{equation}
The above general formula valid in $d$ dimensions is novel. It is always negative since the quantity in square brackets is always positive. For $d=3$ it reduces to $E_{d=3}^{(1)}=-\tfrac{\hbar^2\omega^2}{8\,m\,c^2}\big[6\,n^2 + l^2 +6\,n\,l + 9\,n + 4\,l + \tfrac{15}{4}\big]$ in agreement with previous results \cite{Znojil, Poszwa}. For $d=1$, recall that $l=0$ and $n=N/2$ with $N$ a non-negative integer so that $E_{d=1}^{(1)}=-\tfrac{\hbar^2\omega^2}{32\,m\,c^2}\big[6\,N^2 + 6\,N +3\big]$ in agreement with therelativistic correction to the usual ($d=1$) harmonic oscillator \cite{Griffith, Poszwa}. Our result for $d=2$ is novel and is given by
\beq
\boxed{E_{d=2}^{(1)}=-\frac{\hbar^2\omega^2}{8\,m\,c^2}\big[6\,n^2 + l^2 +6\,n\,l + 6\,n + 3\,l + 2\big]} \,.
\eeq{Ed2}
We also derive the first-order (and second-order) results for the particular case of $d=2$ in a completely different way using ladder operators in polar coordinates in section 4. The results agree providing confirmation of our $d=2$ result.  

\subsection{First-order correction: method II}\label{1RC2}

We now obtain the first-order correction in a different way using a recurrence relation obeyed by the eigenfunctions. Generalized Laguerre polynomials have some nice properties, two of which are \cite{Arfken}
 \begin{equation}
			\label{glpp_1}
			L_n^{(\alpha)}(x) = L_n^{(\alpha+1)}(x) - L_{n-1}^{(\alpha+1)}(x)
		\end{equation}
		and 
	\begin{equation}
			\label{glpp_2}
			x\,L_n^{(\alpha+1)}(x) = (n+\alpha)\,L_{n-1}^{(\alpha)}(x) -(n-x)\,L_n^{(\alpha)}(x)\,.
		\end{equation}
Multiplying the eigenfunction given by \reff{ueta} by $\eta$ and using \reff{glpp_1} and \reff{glpp_2} with $\alpha = l + \frac{d}{2} -1$ and $x=\eta$ we obtain the following recurrence relation
\begin{equation}
	\label{e_wave}
	\eta \,u_{n,l}(\eta) = D_{n,l}\,u_{n+1,l}(\eta) + \frac{E_{n,l}}{\hbar\omega}\,u_{n,l}(\eta) + D_{n-1,l}\,u_{n-1,l}(\eta),
\end{equation}
where 
\beq
D_{n,l} = -\sqrt{(n+1)\left(n+l+\frac{d}{2}\right)}
\eeq{Dnl}
and we used the fact that $A_{nl}$ given by \reff{A} is equal to $\sqrt{\frac{n}{n+l+\frac{d}{2}-1}}A_{n-1 \,l}$.
Multiplying the left-hand side of \reff{e_wave} by $\eta$ and then applying the recurrence relation \reff{e_wave} to each function on the right-hand side yields   
\begin{equation}
	\label{e2_wave}
	\begin{split}
	\eta^2\,u_{n,l}(\eta) =& D_{n,l}\,D_{n+1,l}\,u_{n+2,l} + \frac{D_{n,l}}{\hbar\omega}\,\left[E_{n,l}+E_{n+1,l}\right]\,u_{n+1,l}
	+\left[D_{n,l}^2+\frac{E_{n,l}^2}{\hbar^2\omega^2}+D_{n-1,l}^2\right]\,u_{n,l} 
	\\&+\frac{D_{n-1,l}}{\hbar\omega}\,\left[E_{n-1,l}+E_{n,l}\right]\,u_{n-1,l} +D_{n-1,l}\,D_{n-2,l}\,u_{n-2,l}.
	\end{split}
\end{equation}
The first-order correction, given by \reff{ER1} is then given by
\begin{equation}
	\label{er1_glp}
	E^{(1)} = -\frac{\hbar^2\omega^2}{8mc^2}\left[6n^2+l^2+6nl+3nd+(1+d)l+\frac{d}{4}\Big(2+d\Big)\right]
\end{equation}
where we used \reff{e2_wave}, \reff{Dnl} and \reff{energy}. The above general formula agrees exactly with formula \reff{er1} derived using method I in the previous subsection. This agreement is a strong confirmation of our results.  

\subsection{Splitting of energy levels and degeneracy}\label{Split}

The (unperturbed) energy of the $d$-dimensional isotropic QHO is given by $E= (2n+l+d/2) \hbar\,\omega$. It is convenient to express this as $E=(N+d/2)\,\hbar\,\omega$ where $N=2\,n+l$ has integer values $N=0,1,2,3,...$. Hence, $N=0$ corresponds to the ground state, $N=1$ to the first excited state, etc. The degeneracy of the energy for a given $N$ is well known and given by the binomial coefficient   
\beq
g(N,d)= \binom{N+d-1}{d-1} \,.
\eeq{degen}
For example, the degeneracy is $1$ for $d=1$, $N+1$ for $d=2$ and $(N+1)(N+2)/2$ for $d=3$. The first-order relativistic correction \reff{er1} is always negative and therefore shifts the energy downwards\footnote{The second-order correction, which we calculate in section 5, is positive but much smaller. We do not include it here.}. It also splits the original energy level into a number of distinct levels, each with their own degeneracy.  

We now determine the number of distinct levels for a given $N$ and the degeneracy of each level. We rewrite the first-order relativistic correction \reff{er1} in the form
\beq
E^{(1)}=-\frac{\hbar^2\omega^2}{8\,m\,c^2}\Bigg[\,\frac{3}{2}\,\Big(\dfrac{E}{\hbar\,\omega}\Big)^2 -\frac{1}{2}\,l(l+d-2) +\frac{4\,d-d^2}{8} \Bigg]\,.
\eeq{E1E}
For a given original energy $E$ (hence, a given $N$), the above formula yields a different value for each different value of $l$. In other words, the original energy splits into distinct levels, one for each possible value of $l$. So the number of distinct levels is simply the number of possible $l$ values for a given $N$. $N$ is given by $2\,n+l$. If $N$ is even, $l$ can be equal to $0,2,4,...,N$ so that there are $N/2+1$ possible values. If $N$ is odd, $l$ can be equal to $1,3,5,...,N$ so that there are $(N+1)/2$ possible values. So the original energy level splits into $[N/2 +1]$ distinct levels where $[x]$ is the greatest integer equal to or less than $x$. Therefore $N=\{0,1\}$ has one level (no splitting), $N=\{2,3\}$ splits into two levels, $N=\{4,5\}$ splits into three levels, etc. Note that for the special case of $d=1$, there is no splitting since $l$ is always equal to zero regardless of the value of $N$. The degeneracy of each distinct level is equal to the number of different eigenfunctions of the $L^2$ operator in $d$ dimensions for a given $l$. This is given by \cite{Dong}
\beq
h(l,d)= \dfrac{(2\,l+d-2)(l+d-3)!}{(d-2)!\,l!}\,.
\eeq{hld}
For $d=3$, the above reduces to the expected expression $h=2\,l+1$. An energy level diagram (not to scale) showing the first few energy levels and their splitting is shown in figure 1. The shift in energy is always negative (downwards) but less negative the larger the value of $l$. Note that the degeneracy \reff{degen} of the original unperturbed energy level labeled by $N$ is equal to the sum of the degeneracy \reff{hld} over all the possible $l$ values of the distinct levels the original energy has split into.   

\begin{figure}
 \caption{\label{Energy Levels} Energy level diagram. The unperturbed energy levels labeled by $N$ are shifted downwards in all cases and split into $[N/2+1]$ distinct levels labeled by the value of $l$. The shift is smaller (less negative) for larger $l$. The diagram applies to all dimensions except $d=1$. The degeneracy of each level labeled by $l$ is $h(l,d)$ given by \reff{hld}.}
 \includegraphics{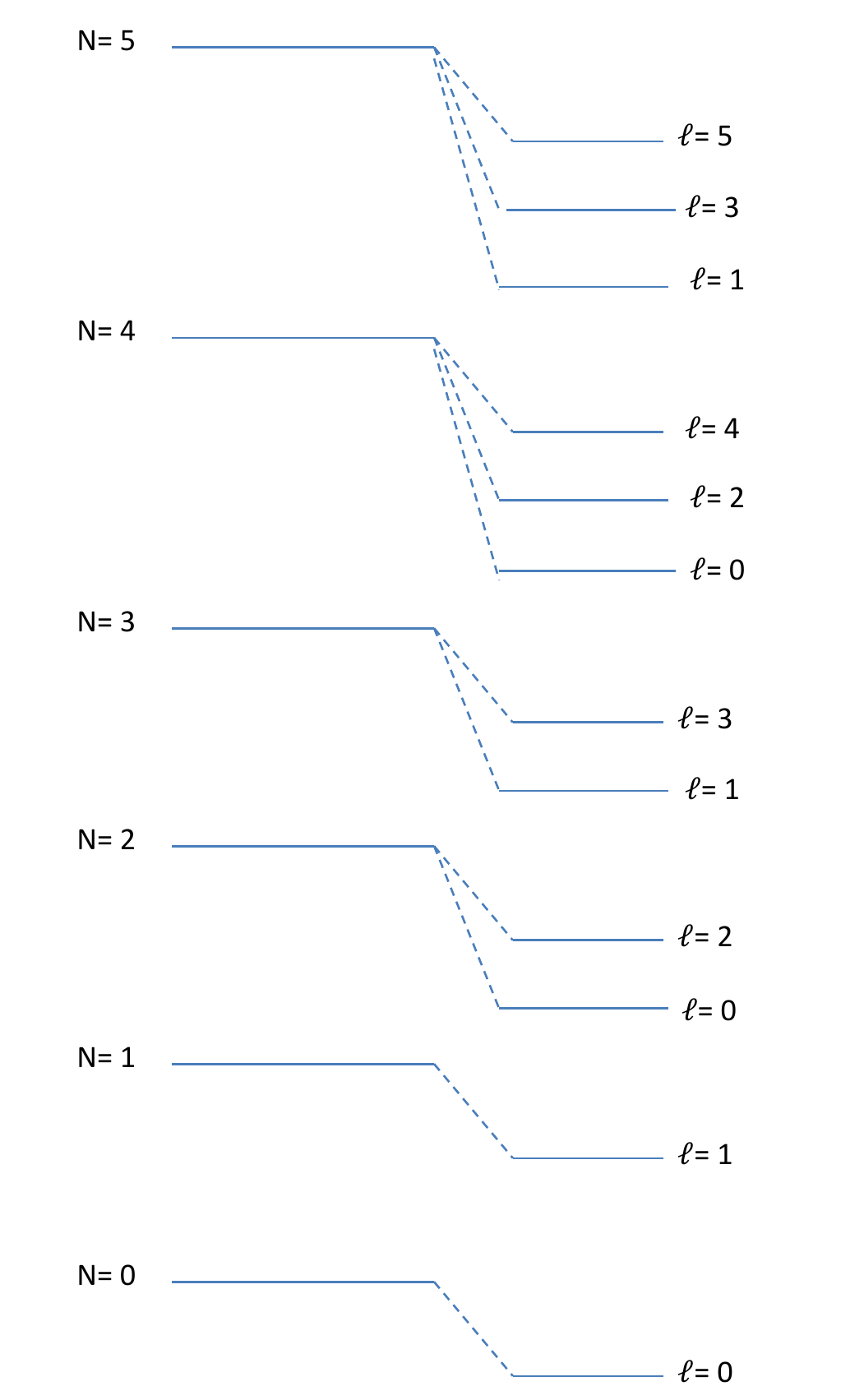}
 \end{figure}
\pagebreak
                
\section{Two-dimensional results using ladder operators in polar coordinates}\label{2DLad}

One is familiar with the use of ladder operators for obtaining the energy levels of the harmonic oscillator in one dimension. In Cartesian coordinates, it is straightforward to generalize this procedure to higher dimensions since the $d$-dimensional oscillator can be thought of as a set of $d$ one-dimensional oscillators, each with their own ladder operators. However, as mentioned in the introduction, if one wants to use non-degenerate perturbation theory, one must use spherical coordinates and the ladder operator technique in $d$-dimensions becomes a bit more cumbersome to use. However, in two dimensions the situation is simple enough that one can define ladder operators in polar coordinates and use them for perturbative calculations. This is an interesting and independent method for obtaining the first and second-order relativistic corrections for the two-dimensional case. 

In appendix C, the ladder operators suited for polar coordinates, labeled $\hat{a}$ and $\hat{b}$, are defined and used to obtain the energy and angular momentum. The eigenstates are labeled $\ket{Nm}$ where $N$ and $m$ are quantum numbers associated with the energy and angular momentum respectively. In particular, in appendix C, the result from acting the ladder operators on the eigenstates are obtained. These results are then used below to calculate the first-order and second-order relativistic corrections.  

It is well known that a charged particle moving in a uniform magnetic field has quantized energies -- called Landau levels -- equivalent to those of the one-dimensional QHO. Examples of real physical systems that exhibit quantized energies equivalent to those of the two-dimensional $isotropic$ QHO appear to be less known. At the end of this section, we discuss briefly such an example. 

\subsection{First-order correction via ladder operators}\label{2DLad1}

The first-order relativistic correction is given by $\frac{-1}{8 m^3 c^2}\bra{Nm}p^4\ket{Nm}$ where $p^2=p_x^2 +p_y^2$. To evaluate this we need to express $p^4$ in terms of the $\hat{a}$ and $\hat{b}$ ladder operators. From \reff{Ladder1} and \reff{Ladder2} we obtain that $p_x= i \frac{\sqrt{\hbar m \omega}}{2}(\hat{a}^{\dagger}+ \hat{b}^{\dagger}-\hat{a}-\hat{b})$ and $p_y=- \frac{\sqrt{\hbar m \omega}}{2}(\hat{a}^{\dagger} - \hat{b}^{\dagger}+\hat{a}-\hat{b})$. After some algebra one obtains that 
\beq
p^2=\hbar\,m\, \omega (\hat{a}^{\dagger}\hat{a} +\hat{b}^{\dagger}\hat{b}-\hat{a}^{\dagger}\hat{b}^{\dagger}-\hat{a}\hat{b} +1)
\eeq{p2}
where \reff{Comm2} was used. The $p^4$ operator has many terms and though not all are needed to calculate the first order correction, most are required to evaluate the second order correction. Squaring the $p^2$ operator and using \reff{Comm2} again yields  
\beq
\begin{split}
p^4 = \hbar^2\,m^2\,\omega^2 \big[(\hat{a}^{\dagger}\hat{a})^2 + (\hat{b}^{\dagger}\hat{b})^2+ (\hat{a}^{\dagger}\hat{b}^{\dagger})^2+ (\hat{a}\hat{b})^2 + 2 +4\hat{a}^{\dagger}\hat{a}\hat{b}^{\dagger}\hat{b} -2\hat{a}^{\dagger}\hat{a}\hat{a}^{\dagger}\hat{b}^{\dagger} \\-2\hat{a}^{\dagger}\hat{a}\hat{a}\hat{b} +3\hat{a}^{\dagger}\hat{a}-2\hat{b}^{\dagger}\hat{b}\hat{a}^{\dagger}\hat{b}^{\dagger}-2\hat{b}^{\dagger}\hat{b}\hat{a}\hat{b}+3\,\hat{b}^{\dagger}\hat{b}-4\hat{a}\hat{b}\,\big]\,.
\end{split}
\eeq{p4}
It is convenient to organize the above terms into a sum of five operators for later calculations,
\beq
p^4=\hbar^2\,m^2\,\omega^2\,(K_0 + R_4 + L_4 + R_2 + L_2)  
\eeq{p42}
where
\begin{align}
K_0&= (\hat{a}^{\dagger}\hat{a})^2 + (\hat{b}^{\dagger}\hat{b})^2+ 4\,\hat{a}^{\dagger}\hat{a}\hat{b}^{\dagger}\hat{b} +3\hat{a}^{\dagger}\hat{a}+3\hat{b}^{\dagger}\hat{b} +2 \quad;\quad R_4= (\hat{a}^{\dagger}\hat{b}^{\dagger})^2\nonumber\\
L_4&= (\hat{a}\hat{b})^2 \quad;\quad R_2= -2\hat{a}^{\dagger}\hat{a}\hat{a}^{\dagger}\hat{b}^{\dagger}-2\hat{b}^{\dagger}\hat{b}\hat{a}^{\dagger}\hat{b}^{\dagger}\quad;\quad L_2= -2\hat{a}^{\dagger}\hat{a}\hat{a}\hat{b}-2\hat{b}^{\dagger}\hat{b}\hat{a}\hat{b} -4\hat{a}\hat{b}\,.
\label{KRL}
\end{align}
The terms in $K_0$ have an equal number of raising and lowering operators (including constants), $R_4$ has four raising operators, $L_4$ has four lowering operators, $R_2$ has two more raising than lowering operators and $L_2$ has two more lowering than raising operators. As a consequence, when the operators act on the state $\ket{Nm}$, $K_0$ does not change the value of the quantum number $N$, $R_4$ raises it by $4$, $L_4$ lowers it by $4$, $R_2$ raises it by $2$ and $L_2$ lowers it by $2$. All five operators leave the quantum number $m$ unchanged. 
   
To calculate the first-order correction, the only part of $p^4$ in \reff{p42} that yields a non-zero value is $K_0$ given by \reff{KRL}.
Using \reff{nm}, the first order relativistic correction is then given by
\begin{align}
E^{(1)}_{d=2}=\frac{-1}{8\, m^3 \,c^2} \bra{Nm}p^4\ket{Nm}&=\dfrac{-\hbar^2\omega^2}{8 \,m \,c^2}\bra{Nm}K_0\ket{Nm}\nonumber\\& =\dfrac{-\hbar^2\,\omega^2}{16 \,m \,c^2} \,(3\,N^2 +6\,N-m^2 +4)\,.
\label{Er1}
\end{align}

The above result matches \reff{Ed2} which was obtained by substituting $d=2$ into our general $d$-dimensional formula \reff{er1} for the first-order relativistic correction. However, \reff{Ed2} is expressed using quantum numbers $n$ and $l$ whereas \reff{Er1} is expressed using quantum numbers $N$ and $m$. These quantum numbers are related in a straightforward fashion. In section \ref{Split} we saw that $N=2n+l$. In two dimensions, the only angular momentum operator is $L_z$ so that $L^2=L_z^2$ and therefore $m^2=l^2$ (note that $l$ is a non-negative integer whereas $m$ here can be positive or negative i.e. $m=\pm \,l$). Substituting $N=2n+l$ and $m^2= l^2$ into \reff{Er1} yields 
\beq
\boxed{E^{(1)}_{d=2} =\dfrac{-\hbar^2\,\omega^2}{16 \,m \,c^2} \,(3\,N^2 +6\,N-m^2 +4)= -\frac{\hbar^2\omega^2}{8\,m\,c^2}\big[6\,n^2 + l^2 +6\,n\,l + 6\,n + 3\,l + 2\big]}
\eeq{2dE1} 
which is exactly the same result as \reff{Ed2}.  
  
\subsection{Second-order correction via ladder operators}\label{2DLad2}

The second order relativistic correction is the sum of two parts, labeled I and II: 
\beq
 E^{(2)}_{d=2} = E^{(2)}_I +E^{(2)}_{II}=\bra{Nm}H_2'\ket{Nm} + \sum_{N'\ne N}\dfrac{|\bra{N'm}H_1'\ket{Nm}|^2}{E_N^0-E_{N'}^0} 
\eeq{Order2}    
where $E^{(2)}_I$ is equal to the first term and $E^{(2)}_{II}$ to the second term. $H_1'$ and $H_2'$ are the relativistic corrections \reff{rel} to the Hamiltonian and $E_N^0=(N+1)\hbar \omega$ and $E_{N'}^0=(N'+1)\hbar \omega$ are the energies of the non-relativistic two-dimensional isotropic harmonic oscillator ($N$ and $N'$ can take on values of 0, 1, 2, 3,...).  

\subsubsection{Part I of 2nd order correction: $E^{(2)}_{I}$}\label{2DLadA}
The first part of the second-order correction is given by 
\beq
E^{(2)}_{I}=\bra{Nm}H_2'\ket{Nm} =\dfrac{1}{16 m^5 c^4}\bra{Nm}p^6\ket{Nm}\,.
\eeq{p6}
The only terms in $p^6$ that we need to keep are those that have the same number of raising and lowering operators. We label this $p_0^6$ and it is given by 
\beq
p^6_0=(p^2)(p^4)= \hbar^3\,m^3\,\omega^3 \Big((\hat{a}^{\dagger}\hat{a} +\hat{b}^{\dagger}\hat{b}+1)\,K_0 -\hat{a}^{\dagger}\hat{b}^{\dagger}\,L_2-ab\,R_2 \Big)
\eeq{p60}
where we used \reff{p2} and \reff{p42} and $K_0$, $L_2$ and $R_2$ are given by \reff{KRL}. Applying \reff{nm} and evaluating the expectation value of the above three terms in the state $\ket{Nm}$ yields
\beq
E^{(2)}_{I}= \dfrac{\hbar^3\omega^3}{32 \, m^2 \,c^4}\big(\,5\,N^3 +15\,N^2 -3\,m^2 -3\,N\,m^2 +22\,N + 12 \big)\,.
\eeq{E2I}
 
\subsubsection{Part II of 2nd order correction: $E^{(2)}_{II}$}\label{2DLadB}

Part II of the 2nd order correction is given by
\beq
E^{(2)}_{II}=\sum_{N'\ne N}\dfrac{|\bra{N'm}H_1'\ket{Nm}|^2}{E_N^0-E_{N'}^0}= \dfrac{1}{64 m^6 c^4}\sum_{N'\ne N}\dfrac{|\bra{N'm}p^4\ket{Nm}|^2}{E_N^0-E_{N'}^0}\,.
\eeq{H1}
Substituting \reff{p42} for $p^4$ and $E_N^0=(N+1)\hbar \omega$ and $E_{N'}^0=(N'+1)\hbar \omega$ into the above equation yields
\beq
E^{(2)}_{II}= \dfrac{\hbar^3\omega^3}{64 m^2 c^4}\sum_{N'\ne N}\dfrac{|\bra{N'm}R_2+L_2+R_4+L_4\ket{Nm}|^2}{N-N'}\,.
\eeq{H1A}
The operator $K_0$, which is part of $p^4$, is not included above because it makes no contribution; $K_0$ acting on $\ket{Nm}$ returns a state with the same value of $N$ but $N'$ cannot equal $N$. Applying \reff{nm} we obtain    
\begin{align*}
R_2\ket{Nm}&=(-2\,N-4)\sqrt{\dfrac{N-m+2}{2}}\sqrt{\dfrac{N+m+2}{2}}\ket{N+2 \;m}\,;
\\
L_2\ket{Nm}&= (-2N)\sqrt{\dfrac{N-m}{2}}\sqrt{\dfrac{N+m}{2}}\ket{N-2 \;m} \,;
\\
R_4\ket{Nm}&=\sqrt{\dfrac{N-m+4}{2}}\sqrt{\dfrac{N+m+4}{2}}\sqrt{\dfrac{N-m+2}{2}}\sqrt{\dfrac{N+m+2}{2}}\ket{N+4 \;m}\,;
\\
L_4\ket{Nm}&=\sqrt{\dfrac{N-m-2}{2}}\sqrt{\dfrac{N+m-2}{2}}\sqrt{\dfrac{N-m}{2}}\sqrt{\dfrac{N+m}{2}}\ket{N-4 \;m}  
\label{coff}
\end{align*}
Substituting the square of the coefficients above into \reff{H1A}, part II of the second-order correction is given by
\beq
E^{(2)}_{II}= \dfrac{\hbar^3\omega^3}{256\, m^2 \,c^4}\big(-17\,N^3 -51\,N^2 +9 \,N\,m^2 -70\,N +9 \,m^2 -36\big)\,.
\eeq{E2II}

The final result $E^{(2)}_{d=2}$ for the second-order correction in two dimensions is given by the sum of the two parts $E^{(2)}_{I}$ given by \reff{E2I} and $E^{(2)}_{II}$ given by \reff{E2II}:
\beq
E^{(2)}_{d=2}= E^{(2)}_{I} + E^{(2)}_{II}=\dfrac{\hbar^3\,\omega^3}{256\, m^2 \,c^4}\big(23\,N^3 +69\,N^2 -15 \,N\,m^2 +106\,N -15 \,m^2 +60\big)\,.
\eeq{E2F}
In section 5, we obtain the general formula \reff{E2d} for the second-order correction valid in any dimension $d$. When $d=2$ is substituted in \reff{E2d} it yields \reff{E2d2}. We therefore want to compare the above result \reff{E2F} to \reff{E2d2}. As discussed previously (see discussion below \reff{Er1}), for the comparison, we need to replace $N$ by $2n+l$ and $m^2$ by $l^2$ in \reff{E2F}. This yields the expression 
\beq
\boxed{
\begin{split}
E^{(2)}_{d=2}&=\dfrac{\hbar^3\,\omega^3}{256\, m^2 \,c^4}\big(184\,n^3 +276\, n^2l +108\,nl^2 +8\,l^3 \\&\qquad\qquad\qquad+276\,n^2 +276\,nl +54\,l^2+212\,n +106\,l +60)
\end{split}}
\eeq{E2G}
which is in agreement with \reff{E2d2}. This provides a strong confirmation of the ladder operator technique used here and a cross-check between the different methods.

\subsection{Physical system with quantized energies equal to those of the two-dimensional isotropic QHO}\label{LGP} 

We discuss briefly here an example of a non-relativistic physical system that has quantized energy levels equivalent to those of the non-relativistic two-dimensional $isotropic$ QHO.         

A particle of charge $q$ and mass $m$ moving non-relativistically in the $x\!-\!y$ plane under a uniform magnetic field $\textbf{B}= B_0 \,\hat{z}$ where $B_0$ is a constant, have quantized energies given by $E_n= (n+ \frac{1}{2})\, \hbar\,\omega_c$ where $n$ is a non-negative integer and $\omega_c= |q\,B_0|/m$ is referred to as the cyclotron frequency. These are the well known Landau levels and they are identical to those of the one-dimensional QHO but in contrast are continuously degenerate (see \cite{Landau, Ballentine} for an introduction). Note that though the particle's motion is constrained to two dimensions, the energies are those of the one-dimensional oscillator. The particle is of course free to move in the $z$-direction but this would merely add the usual kinetic energy of motion $p_z^2/(2\,m)$ to the Landau levels. We also do not include here the Zeeman splitting due to the coupling of the spin to the magnetic field.   

Consider now adding to the above scenario a linear electric field $\textbf{E}=-k \,z\,\hat{z}$ where $k$ is a constant. This leads to oscillations in the $z$-direction with angular frequency $\omega_1=\sqrt{\frac{q\,k}{m}}$ ($q\,k$ is positive, so that $q$ and $k$ are both positive or both negative). The $z$ and $x\!-\!y$ motion are independent of each other (i.e. one can express the Hamiltonian as $H=H_{xy} + H_z$ where $H_{xy}$ and $H_z$ commute). The quantized energies are then given by the sum of the individual oscillator energies:
\beq 
E= (n_1+ 1/2) \,\hbar\omega_1 + (n_2 +1/2) \,\hbar\omega_c 
\eeq{EB}
where $n_1, n_2$ are non-negative integers. If we now choose the magnetic field strength to be $B_0=\sqrt{\frac{m\,k}{q}}$ then $\omega_1=\omega_c=\omega$ and we obtain quantized energies equal to those of the non-relativistic two-dimensional isotropic QHO,
\beq
E= (n_1+ n_2 +1)\,\hbar\omega = (N+1)\,\hbar\,\omega
\eeq{2DH}
where $N=n_1+n_2$ is any non-negative integer and $\omega=\sqrt{\frac{q\,k}{m}}$. Therefore, in principle, one should be able to construct a physical system with a linear electric field and a constant magnetic field that has the same quantized energies as the non-relativistic two-dimensional isotropic QHO. In the conclusion, we discuss how the relativistic corrections to this physical system would be of interest to study in the future.

\section{Second-order relativistic correction: general formula}

We now calculate the second-order relativistic correction $E^{(2)}$ and obtain a general formula valid in $d$ dimensions. This correction is the sum of two parts, labeled I and II: 
\beq
 E^{(2)} = E^{(2)}_I +E^{(2)}_{II}= \langle \psi_n^0 |H_2'| \psi_n^0\rangle + \sum_{m\neq n} \frac{\abs{\langle \psi_m^0 |H_1'| \psi_n^0\rangle}^2}{E_n^0-E_m^0}, 
\eeq{Order3} 
where $E^{(2)}_I$ is equal to the first term and $E^{(2)}_{II}$ to the second term. $H_1'$ and and $H_2'$ are the relativistic corrections \reff{rel} to the Hamiltonian and $E_n^0$ and $E_{m}^0$ are the energies of the unperturbed or non-relativistic isotropic QHO in state $\psi_n^0$ and $\psi_m^0$ respectively. From the spectral point of view, the Hamiltonian operator \reff{Intro} and the operator $\sqrt{m^2\,c^4 + m^2\,\omega^2\,r^2\,c^2} -m\,c^2 + \frac{p^2}{2\,m}$ are equivalent \cite{Lucha1, Znojil}. We can therefore express $H_1'$ and and $H_2'$ in terms of $r$ (or $\eta=\frac{m\omega\,r^2}{\hbar}$) by replacing $p^2$ by $m^2\,\omega^2\,r^2$ i.e. $H_1'=-\frac{m\,\omega^4\,r^4}{8\,c^2}=-\frac{\hbar^2\,\omega^2\,\,\eta^2}{8\,m\,c^2}$ and $H_2'= \frac{m\,\omega^6\,r^6}{16\,c^4}=\frac{\hbar^3\,\omega^3\,\eta^3}{16\,m^2c^4}$. These expressions for $H_1'$ and $H_2'$ were used in \cite{Znojil,Poszwa}.

\subsection{Part I of second-order correction}

Using \reff{e_wave} and \reff{e2_wave} the expectation value of $\eta^3$ is 
\begin{align*}
	\langle \eta^3\rangle=&\left[\frac{D_{n,l}^2}{\hbar\omega}\left(E_{n+1,l}+E_{n,l}\right)+\frac{E_{n,l}}{\hbar\omega}\left(D_{n,l}^2
	+\frac{E_{n,l}^2}{\hbar^2\omega^2}+D_{n-1,l}^2\right)+\frac{D_{n-1,l}^2}{\hbar\omega}\left(E_{n-1,l}+E_{n,l}\right)\right]\,.
\end{align*}
Substituting  \reff{Enl} and \reff{Dnl} for $E_{n,l}$ and $D_{n,l}$ respectively yields  
\begin{equation}
	\label{E21d}
	\begin{split}
	E^{(2)}_I &= \frac{\hbar^3\,\omega^3}{16\,m^2\,c^4}\,\langle\eta^3\rangle\\&=\frac{\hbar^3\,\omega^3}{16\,m^2\,c^4}\,\bigg[20\,n^3+15\,n^2\,d+\left(4+3d+3d^2\right)\,n+l^3+\frac{3}{2}\,\left(d+2\right)\,l^2+\left(2+3d+\frac{3}{4}d^2\right)\,l\\&
	\quad\quad+30\,n^2\,l+12\,n\,l^2+6\left(1+2d\right)n\,l+\frac{d}{8}\left(8+6d+d^2\right)\bigg]\,.
	\end{split}
\end{equation}
\subsection{Part II of second-order correction }

The second part of the 2nd-order correction is given by
\begin{equation}
	E^{(2)}_{II} = \sum_{m\neq n} \frac{\abs{\langle \psi_m^0 | H_1'| \psi_n^0\rangle}^2}{E_n^0-E_m^0}\,.
\end{equation}
Substituting $H_1'=-\frac{\hbar^2\,\omega^2\,\,\eta^2}{8\,m\,c^2}$ and the energy of the unperturbed oscillator \reff{Enl} into the above yields 
\begin{align*}
	E^{(2)}_{II} = \frac{\hbar^3\,\omega^3}{128\,\,m^2\,c^4}\sum_{m\neq n} \frac{\abs{\langle u_{m,l}| \eta^2|u_{n,l}\rangle}^2}{n-m}\,.
\end{align*}
The above can be evaluated using \reff{e2_wave}:  
\begin{align}
E^{(2)}_{II} &=\frac{\hbar^3\,\omega^3}{128\,m^2\,c^4}\sum_{m\neq n}\Bigg[ \frac{\abs{\langle u_{m,l}| D_{n,l}D_{n+1,l}u_{n+2,l}
	\rangle}^2}{n-m} 
	+\frac{\abs{\langle u_{m,l}|D_{n-1,l}D_{n-2,l}u_{n-2,l}\rangle}^2}{n-m}\nonumber
	\\&\qquad+\frac{\abs{\left\langle u_{m,l}|\frac{D_{n,l}}{\hbar\omega}(E_{n,l}+E_{n+1,l})u_{n+1,l}\right\rangle}^2}{n-m}
	+\frac{\abs{\left\langle u_{m,l}| \frac{D_{n-1,l}}{\hbar\omega}(E_{n,l}+E_{n-1,l})u_{n-1,l}\right\rangle}^2}{n-m}\Bigg].
\label{EII2}
\end{align}
Substituting \reff{Enl} and \reff{Dnl} for $E_{n,l}$ and $D_{n,l}$ respectively into \reff{EII2} yields the expression for part II of the second-order correction:
\begin{equation}
	\label{E22d}
	\begin{split}
	E^{(2)}_{II} =& -\frac{\hbar^3\omega^3}{256m^2c^4}\Big[
	136n^3+102n^2d+(21d^2+18d+20)n+8l^3+(12d+18)l^2\\&+(6d^2+18d+10)l+204n^2l+84nl^2+(84d+36)nl
	+\frac{1}{2}(2d^2+9d+10)d
	\Big]\,.
	\end{split}
\end{equation}
The final expression for the full second-order relativistic correction, the sum of part I given by \reff{E21d} and part II given by \reff{E22d}, is
\begin{tcolorbox}[standard jigsaw,boxrule=0.5pt,opacityback=0, width=6in, height=1.5in]
	\begin{align}
	E^{(2)} &= \frac{\hbar^3\omega^3}{256\,m^2c^4}\Bigg[
	184\,n^3+138\,n^2d+(27\,d^2+30\,d+44)n+8\,l^3+(12\,d+30)\,l^2\nonumber
	\\&+(6\,d^2+30\,d+22)\,l+276\,n^2l+108\,nl^2+(108\,d+60)\,nl+\left(d^2+\frac{15}{2}d+11\right)d\,
	\Bigg].
	\label{E2d}
	\end{align}
	\end{tcolorbox}
Note that the above second-order correction is positive and valid for any dimension $d$. It is novel and reduces to previous results \cite{Poszwa} for $d=1$ and $d=3$ respectively. In one dimension recall that we set $l=0$ and $n= N/2$ where $N$ is a non-negative integer and this yields $E_{d=1}^{(2)} = \frac{\hbar^3\omega^3}{512m^2c^4}\left(46N^3+69N^2+101N+39\right)$. In three dimensions, one obtains $E_{d=3}^{(2)} = \frac{\hbar^3\omega^3}{256m^2c4}\Big[184n^3+414n^2+377n+8l^3+66l^2+166l+276n^2l+108nl^2+330nl+\frac{255}{2}\Big]$. Both results are in agreement with \cite{Poszwa}. 

The two-dimensional result is novel and given by
	\beq
	E^{(2)}_{d=2} \!=\! \frac{\hbar^3\omega^3}{256\,m^2c^4}\Big[
	184\,n^3 \!+\! 276\,n^2 \!+\! 212\,n \!+ \!8\,l^3 \!+\! 54\,l^2
	\!+\! 106\,l \!+\! 276\,n^2l \!+\! 108\,nl^2 \!+\! 276\,nl \!+ 60\,
	\Big].
	\eeq{E2d2}
Note that the above result for $d=2$ is in agreement with the second-order correction \reff{E2G} obtained in the previous section via the ladder operator technique. The fact that the $d=1$ and $d=3$ results agree with previous work and that the $d=2$ results agree with those we obtained via the independent ladder operator technique is a strong confirmation of our general formula 
\reff{E2d}.  

\section{Conclusion}

In this paper we considered the relativistic Hamiltonian operator given by \reff{Intro}. This can be viewed as representing a relativistic harmonic oscillator obeying the spinless Salpeter equation. This cannot be solved exactly analytically. We used perturbation theory to obtain the general formulas \reff{er1} and \reff{E2d} for the first-order and second-order relativistic corrections respectively. They are expressed in terms of the non-negative integral quantum numbers $n$ and $l$ and are valid in any dimension $d$. Both formulas reduce to previously known results \cite{Znojil,Poszwa} for the cases of $d=1$ and $d=3$. We saw that the the relativistic corrections split the original energy, labeled by $N$, into [N/2 +1] distinct levels (labeled by $l$) each with their own degeneracy given  
by \reff{hld}. The results for $d=2$ are novel and match those that we obtained using an independent ladder operator technique (equation \reff{2dE1} for the first-order and equation \reff{E2G} for the second-order). 

In all, we used three independent methods in this paper. One method was a $d$-dimensional Kramers-Pasternak type relation that we derived in appendix B which allows one to evaluate the expectation value of a certain power of $r$ in terms of the expectation values of other powers of $r$. This was used to obtain the general formula for the first-order correction (method I). The second method made use of a recurrence relation for the eigenfunctions when they are multiplied by $\eta$, $\eta^2$ or $\eta^3$ where $\eta$ is proportional to $r^2$. This was used to obtain the general formula for the first-order correction in an alternative fashion (method II) and to obtain the general formula for the second-order correction. The third method made use of ladder operators in polar coordinates to evaluate the first and second-order corrections for the two-dimensional case. The results of all three methods agree with each other. Relativistic wave equations reduce to Schr\"odinger's equation plus corrections in the non-relativistic limit \cite{Maggiore}. These corrections, besides others, will typically include the kinematic $p^4$ and $p^6$ perturbations, so that in most cases, the different techniques used here should be very useful.  

It is well known that a charged particle moving non-relativistically in a uniform magnetic field has discrete energies -- called Landau levels -- that are  equivalent to those of the one-dimensional QHO. We pointed out that a charged particle moving non-relativistically in a uniform magnetic field $\textbf{B}= B_0 \,\hat{z}$ and a linear electric field $\textbf{E}=-k \,z\,\hat{z}$, where $B_0$ and $k$ are constants, can have discrete energies that are equivalent to those of the two-dimensional $isotropic$ QHO as long as the magnetic field is of a particular strength. What are the relativistic energies of this system? There is no exact analytical solution to the Dirac equation for this system\footnote{Known exact solutions to the Dirac equation are limited. The following cases have been solved exactly: a constant magnetic field, a constant electric field, a constant orthogonal electric and magnetic field,  the Coulomb potential and a few others (see \cite{Johannes} and references therein).}. One can either solve it exactly numerically or obtain an analytical approximation using perturbation theory. Though the non-relativistic energies of the charged particle in the physical scenario discussed above match those of the two-dimensional isotropic QHO, there is no reason that the relativistic corrections should match exactly (besides the differences due to the coupling of the spin with the magnetic field). However, it is expected that they should resemble them quite closely in form.

\begin{appendices}
\numberwithin{equation}{section}
\setcounter{equation}{0}
\section{Radial equation in $d$ dimensions}

In $d$ spatial dimensions, Schr\"odinger's equation is given by
\begin{equation}
	\label{Schrod}
	-\frac{\hbar^2}{2m}\Delta \,\psi +V\,\psi = E\,\psi,
\end{equation}
where $\Delta$ is the Laplacian operator, which in spherical coordinates is
\begin{equation}
	\label{Laplacian}
	\Delta = \frac{1}{r^{d-1}}\frac{\partial}{\partial r}\left(r^{d-1}\frac{\partial}{\partial r}\right)+\frac{1}{r^2}\,\Delta_{s^{d-1}}.
\end{equation}
Here, $\Delta_{s^{d-1}}$ is the Laplace-Beltrami operator on the $(d-1)$-sphere and it only contains derivatives with respect to the angles $\theta$, $\phi_1$, $\phi_2$, ... , $\phi_{d-2}$.

We now assume that the potential is spherically symmetric so that $V=V(r)$ and look for solutions of Schr\"odinger's equation in separable form i.e.
$$\psi = R(r)\,\Omega(\theta,\phi_1,...,\phi_{d-2})=R\,\Omega.$$
Then \reff{Schrod} take the form
$$R\,\Delta_{s^{d-1}}\Omega - \frac{2m}{\hbar^2}\,r^2\,\big(V(r)-E\big)\,R\,\Omega+\frac{\Omega}{r^{d-3}}\frac{\partial}{\partial r}\left(r^{d-1}\frac{\partial R}{\partial r}\right)=0.$$
Dividing both sides by $\psi = R\,\Omega$, one obtains
$$\frac{\Delta_{s^{d-1}}\Omega}{\Omega} +\Big[-\frac{2m\,r^2}{\hbar^2}\,\big(V(r)-E\big)+\frac{1}{R\,r^{d-3}}\frac{\partial}{\partial r}\left(r^{d-1}\frac{\partial R}{\partial r}\right)\Big]=0.$$
The first term depends only on angles and the second term in square brackets depends only on $r$. Therefore each term must be equal to a constant. The constant is $l\,(l+d-2)$ where $\ell$ is a non-negative integer and this can be obtained by solving the angular part of the equation \cite{Dong} (in three dimensions it reduces to the familiar value of $\ell(\ell+1)$). Equating the radial part in square brackets to this constant yields the equation
\begin{equation}
	\label{rad}
	-\frac{2mr^2}{\hbar^2}\big(V(r)-E\big)+\frac{1}{R\,r^{d-3}}\frac{\partial}{\partial r}\left(r^{d-1}\frac{\partial R}{\partial r}\right)=l(l+d-2).
\end{equation}
Let $u(r) = r\,R(r)$. Then   
\begin{align*}
	\frac{d}{dr}\left(r^{d-1}\frac{dR}{dr}\right) =-(d-3)\,u\,r^{d-4}+(d-3)\,u'r^{d-3}+u''\,r^{d-2}
\end{align*}
where a prime denotes derivative with respect to $r$. Expressing \reff{rad} in terms of $u$ yields the radial equation in $d$ dimensions:
\begin{equation}
	\label{radial_equation}
	u''=\left[\frac{2m}{\hbar^2}\big(V(r)-E\big)+\frac{\big(d-3+l(l+d-2)\big)}{r^2}\right]\,u - \frac{(d-3)}{r}\,u'
\end{equation}

\section{Derivation of $d$-dimensional Kramers-Pasternak type relation}

We now derive a $d$ dimensional Kramers-Pasternak type relation. Multiplying \reff{sandwich} by $u\,r^s\,r^{d-3}$ and then integrating yields 
\begin{align}
	\int u\,r^s\,u''\, r^{d-3}\,dr &= \frac{m^2\omega^2}{\hbar^2}\,\langle\, r^{s+2}\,\rangle -\frac{2mE}{\hbar^2}\,\langle\, r^s\,\rangle +\left[d-3+l(l+d-2)\right]\,\langle\, r^{s-2}\,\rangle \nonumber\\&\qquad\qquad- (d-3)\int u\,r^{s-1}\,u'\,r^{d-3}\,dr.
\label{upp}
\end{align}
where $\langle\, f(r)\,\rangle$ denotes the expectation value of $f(r)$ defined as $\int u\,f(r)\,u \,r^{d-3}\, dr$. Integrating by parts we obtain 
\beq
\int u\,r^s\,u''\,r^{d-3}\, dr =-\int u'\,r^s \,u'\,r^{d-3}\, dr + (3-d-s)\,\int u\,r^{s-1}\,u'\,r^{d-3} \,dr 
\eeq{Urs}
where we used the fact that the boundary term is equal to zero. Integrating by parts again, one obtains that
\beq
\int u \,r^s \,u' \,r^{d-3}\,dr = \dfrac{(3-d-s)}{2} \,\langle\, r^{s-1}\,\rangle
\eeq{uprime}
and 
\beq
\int u' \,r^s \,u' \,r^{d-3}\,dr = \dfrac{2}{(2-d-s)} \,\int u'' \,r^{s+1} \,u' \,r^{d-3}\,dr 
\eeq{u2prime}
Substituting $u''$ from \reff{sandwich} into \reff{u2prime} and then using the result \reff{uprime} yields
\begin{multline}
\int u' \,r^s \,u' \,r^{d-3}\,dr =\dfrac{1}{4-d+s}\Big[\frac{m^2\omega^2}{\hbar^2}\,(d+s)\,\langle\, r^{s+2}\,\rangle+\frac{2\,m\,E}{\hbar^2} \,(2-d-s)\,\langle\, r^{s}\,\rangle\\+\big(d-3+l\,(l+d-2)\big)\,(s+d-4)\,\langle\, r^{s-2}\,\rangle\Big]\,.
\label{u3prime}
\end{multline}
Substituting \reff{u3prime} and the result \reff{uprime} into \reff{Urs} and then substituting that result into \reff{upp} yields a $d$-dimensional Kramers-Pasternak type relation:
\begin{multline}
\frac{m^2\omega^2}{\hbar^2}\,(2s+4)\langle\, r^{s+2}\,\rangle -\frac{2mE}{\hbar^2} \,(2s+2)\,\langle\, r^{s}\,\rangle
\\+\bigg[\big(d-3+l\,(l+d-2)\big)\,2s + \dfrac{s}{2}(4-d-s)(4-d+s)\bigg]\langle\, r^{s-2}\,\rangle
=0\,.
\label{dKramers}
\end{multline}
The above equation is used in section 3.1 to evaluate $\langle\, r^4\,\rangle$.

\section{Ladder operators in polar coordinates}

The ladder operators in polar coordinates are defined in terms of those in Cartesian coordinates so it will be beneficial to quickly review the Cartesian case as the results then transfer readily to polar coordinates. In two dimensions, the ladder operators in Cartesian coordinates are defined as:
\beq
\hat{a}_x= \dfrac{1}{\sqrt{2\hbar\, m \,\omega}}\,(m\,\omega\,\hat{x}+i\hat{p}_x) \word{and} \hat{a}_y=\dfrac{1}{\sqrt{2\hbar\, m \omega}}\,(m\omega\,\hat{y}+i\hat{p}_y)
\eeq{Ladder1}
and obey the commutation relations 
\beq
[\hat{a}_x,\hat{a}_x^{\dagger}]=1 \quad;\quad [\hat{a}_y,\hat{a}_y^{\dagger}]=1 \word{with all other commutations being zero}.
\eeq{Comm1}
Here $\hat{a}_x^{\dagger}=\tfrac{1}{\sqrt{2\hbar\, m \,\omega}}\,(m\,\omega\,\hat{x}-i\hat{p}_x)$ is the conjugate transpose of $\hat{a}_x$. The unperturbed (non-relativistic) Hamiltonian operator is given by 
\beq
H_0= \dfrac{\hat{p}_x^2}{2m} +\dfrac{\hat{p}_y^2}{2m} + \dfrac{1}{2} m \,\omega^2 (\hat{x}^2 +\hat{y}^2)= (\hat{a}_x^{\dagger}\hat{a}_x +\hat{a}_y^{\dagger}\hat{a}_y +1) \hbar \omega =(\hat{n}_x +\hat{n}_y +1) \hbar \omega =(\hat{N} +1)\hbar \omega
\eeq{H0}
where $\hat{n}_x=\hat{a}_x^{\dagger}\hat{a}_x$, $\hat{n}_y=\hat{a}_y^{\dagger}\hat{a}_y$ and $\hat{N}=\hat{n}_x +\hat{n}_y$. The Hamiltonian breaks up into a sum of two one-dimensional oscillators, one in the $x$ and one in the $y$. The ladder operators have the following properties:
\beq
[\hat{N},\hat{a_x}^{\dagger}]= \hat{a_x}^{\dagger}\quad;\quad [\hat{N},\hat{a_y}^{\dagger}]= \hat{a_y}^{\dagger}\quad;\quad [\hat{N},\hat{a_x}]= -\hat{a_x} \quad;\quad 
[\hat{N},\hat{a_y}]= -\hat{a_y}\,.
\eeq{Ladderxy} 
This implies that when $\hat{a}_x^{\dagger}$ or $\hat{a}_y^{\dagger}$ acts on an eigenstate of $\hat{N}$ it increases its eigenvalue by 1 whereas $\hat{a}_x$ or $\hat{a}_y$ decreases it by 1. For this reason, the former are called raising operators and the latter lowering operators. The harmonic oscillator potential is non-negative and therefore the energy $E$ must also be non-negative. This implies that there is an eigenstate $\psi_0$ (the ground state) of $\hat{n}$ that yields zero when the lowering operators $\hat{a_x}$ or ${a_y}$ act on it i.e. when $\hat{n}_x$ and $\hat{n}_y$ act on $\psi_0$, they yield zero so that their eigenvalues (and those of $\hat{N}$) start at zero and increase by unity (are non-negative integers). Since $[\hat{n}_x,\hat{n}_y]=0$, we can construct simultaneous eigenstates of these two operators (they are then eigenstates of $\hat{N}$ and the Hamiltonian $H_0$). We denote the eigenstates as $\ket{n_x n_y}$, and their respective eigenvalues will be $n_x$ and $n_y$, both non-negative integers. The energy of the two-dimensional oscillator is then $(N+1) \hbar \,\omega$ where $N=n_x+n_y$ is also a non-negative integer. The energy for a given $N$ is then clearly $(N+1)$-fold degenerate.  

We define the ladder operators $\hat{a}$ and $\hat{b}$ as
\beq
\hat{a}=\dfrac{1}{\sqrt{2}}(\hat{a}_x + i \hat{a}_y) \word{and} \hat{b}=\dfrac{1}{\sqrt{2}}(\hat{a}_x - i \hat{a}_y)\,. 
\eeq{Ladder2}
With the above definitions, it follows from \reff{Comm1} that they have the following property:  
\beq
[\hat{a},\hat{a}^{\dagger}]=1 \word{and} [\hat{b},\hat{b}^{\dagger}]=1 \word{(with all other commutators zero)}. 
\eeq{Comm2}
It is convenient to define the operators 
\beq
\hat{n}_a= \hat{a}^{\dagger} \hat{a} \word{and} \hat{n}_b= \hat{b}^{\dagger} \hat{b}\,.
\eeq{nanb}
The operator $\hat{N}=\hat{n}_x+\hat{n}_y$ takes on the similar form   
\beq
\hat{N}= \hat{n}_a + \hat{n}_b
\eeq{nab}
and we obtain commutation relations similar to \reff{Ladderxy}:  
\beq
[\hat{N},\hat{a}^{\dagger}]= \hat{a}^{\dagger}\quad;\quad [\hat{N},\hat{b}^{\dagger}]= \hat{b}^{\dagger}\quad;\quad [\hat{N},\hat{a}]= -\hat{a} \quad;\quad 
[\hat{N},\hat{b}]= -\hat{b}\,.
\eeq{Ladder3}  
Again, this implies that when $\hat{a}^{\dagger}$ and $\hat{b}^{\dagger}$ act on an eigenstate of $\hat{N}$, they increase its eigenvalue by unity whereas $\hat{a}$ and $\hat{b}$ decrease its eigenvalue by unity. 

In two dimensions, the only angular momentum operator is $L_z \equiv \hat{x}\hat{p}_y-\hat{y}\hat{p}_x=-i\hbar\frac{\partial}{\partial \theta}$where $\theta$ is the usual angle in polar coordinates.  We can express $L_z$ in terms of the $\hat{n}_a$ and $\hat{n}_b$ operators: 
\beq
L_z=i\hbar\,(a_y^{\dagger}a_x-a_ya_x^{\dagger})=\hbar \,(\hat{n}_b - \hat{n}_a)
\eeq{Lz}
where \reff{Ladder1},\reff{Comm1},\reff{Ladder2},\reff{Comm2} and \reff{nanb} were used. From \reff{Comm2} we can deduce the following commutation relations:
\beq
[L_z, \hat{a}]=\hbar\, \hat{a}\quad;\quad [L_z, \hat{a}^{\dagger}]=-\hbar \,\hat{a}^{\dagger}\quad;\quad [L_z, \hat{b}]=-\hbar \,\hat{b} \quad;\quad [L_z, \hat{b}^{\dagger}]=\hbar \,\hat{b}^{\dagger} \,.
\eeq{ang}
This implies that when $\hat{a}$ and $\hat{b}^{\dagger}$ act on an eigenstate of $L_z$ they raise the angular momentum by $\hbar$ whereas $\hat{a}^{\dagger}$ and $\hat{b}$ lower it by $\hbar$. Because of this property, the operators $\hat{a}$ and $\hat{b}$ (and their conjugate transpose) are the ladder operators appropriate for use with polar coordinates.  

Note that $[\hat{n}_a,\hat{n}_b]=0$ so we can construct simultaneous eigenstates of the two operators. We label this product state $\ket{n_a \,n_b}$ with $\hat{n}_a\ket{n_a\,n_b}= n_a \ket{n_a\,n_b}$ and $\hat{n}_b\ket{n_a\,n_b}= n_b \ket{n_a\,n_b}$ where both $n_a$ and $n_b$ are non-negative integers (this follows from exactly the same argument we gave above as to why $n_x$ and $n_y$ are non-negative integers). The state  
$\ket{n_a \,n_b}$ are also simultaneous eigenstates of $L_z$ and $\hat{N}$ -- in accord with the fact that $[\hat{N},L_z]=0$ -- with eigenvalues $\hbar(n_b-n_a)$ and $n_a + n_b$ respectively .  It is more convenient to work with the alternate product state $\ket{Nm}$ such that 
\beq
\hat{N}\ket{Nm}=N\ket{Nm} \;\; (N=0,1,2,3,...)\;;\; L_z\ket{Nm}=\hbar\, m \ket{Nm} \;\; (m=-N,-N+2,...,N-2,N)
\eeq{nm2}
where $N=n_a+n_b$ and $m=n_b-n_a$. For a given non-negative integer $N$, $m$ runs from $-N$ to $N$ in steps of two. This follows from the fact that for a given $N$, the allowed pairs $(n_a,n_b)$ are $(N,0),(N-1,1),...,(1,N-1),(0,N)$ which yields $m=-N,-N+2,...,N-2,N$.  The energy is given by $(N+1)\hbar \omega$ and there are $N+1$ values of $m$ for each energy level so that the energies of the two-dimensional oscillator are $(N+1)$-fold degenerate in agreement with the degeneracy previously obtained using Cartesian coordinates.                

For later calculations, we need to determine the coefficients when $\hat{a}$, $\hat{a}^{\dagger}$, $\hat{b}$ and $\hat{b}^{\dagger}$ act on $\ket{Nm}$. From \reff{nab} and \reff{Lz} we know that $\hat{n}_a=\hat{a}^{\dagger}\hat{a}=\frac{1}{2}(\hat{N} -L_z/\hbar)$ and $\hat{n}_b=\hat{b}^{\dagger}\hat{b}=\frac{1}{2}(\hat{N} +L_z/\hbar)$. Therefore $\bra{Nm} \hat{a}^{\dagger}\hat{a}\ket{Nm}= \frac{N-m}{2}$ and $\bra{Nm} \hat{b}^{\dagger}\hat{b}\ket{Nm}= \frac{N+m}{2}$. It then follows from \reff{Comm2} that $\bra{Nm} \hat{a}\hat{a}^{\dagger}\ket{Nm}= \frac{N-m+2}{2}$ and $\bra{Nm} \hat{b}\hat{b}^{\dagger}\ket{Nm}= \frac{N+m+2}{2}$. From these above results and using \reff{Ladder3} and \reff{ang}, we obtain the following:
\begin{align}
&\hat{a}\ket{Nm}= \sqrt{\tfrac{N-m}{2}}\,\ket{N-1\, m+1}  \quad & \hat{b}&\ket{Nm}= \sqrt{\tfrac{N+m}{2}}\,\ket{N-1 \,m-1}\nonumber\\
&\hat{a}^{\dagger}\ket{Nm}= \sqrt{\tfrac{N-m+2}{2}}\,\ket{N+1 \, m-1} \quad  & \hat{b}^{\dagger}&\ket{Nm}= \sqrt{\tfrac{N+m+2}{2}}\,\ket{N+1 \,m+1}\,.\label{nm}
\end{align}

With the above results, we can build the state $\ket{Nm}$ from the ground state $\ket{00}$: 
\beq 
\ket{Nm}=\dfrac{1}{(\frac{N-m}{2})!(\frac{N+m}{2})!}(\hat{a}^{\dagger})^{(N-m)/2}\hat{b}^{\dagger})^{(N+m)/2}\ket{00} \,. 
\eeq{nm3}
\end{appendices}  
\section*{Acknowledgments}
A.E. acknowledges support from a discovery grant of the National Science and Engineering Research Council of Canada (NSERC) and P.L. acknowledges support from NSERC's USRA program.

\end{document}